\documentclass{emulateapj}
\usepackage[backref,breaklinks,colorlinks=true,linkcolor = blue,citecolor=blue]{hyperref}
\usepackage{enumitem}  
\usepackage{graphics}

\setlist[enumerate]{wide=\parindent}

\usepackage{natbib}
\shorttitle{M101 Satellites}
\shortauthors{Danieli et al.}

\begin{document}

%\title{Low luminosity dwarfs around nearby galaxies: I. M101 group}
\title{The Dragonfly Nearby Galaxies Survey. III. The Luminosity Function of the M101 Group}

\author{Shany Danieli\altaffilmark{1,2,3}}
\author{Pieter van Dokkum\altaffilmark{3}}
\author{Allison Merritt\altaffilmark{3}}
\author{Roberto Abraham\altaffilmark{4,5}}
\author{Jielai Zhang\altaffilmark{4,5,6}}
\author{I. D. Karachentsev\altaffilmark{7}}
\author{L. N. Makarova\altaffilmark{7}}

\altaffiltext{1}{Department of Physics, Yale University, New Haven, CT 06520, USA}
\altaffiltext{2}{Yale Center for Astronomy and Astrophysics, Yale University, New Haven, CT 06511, USA}
\altaffiltext{3}{Department of Astronomy, Yale University, New Haven, CT 06511, USA}
\altaffiltext{4}{Department of Astronomy and Astrophysics, University of Toronto, Toronto ON, M5S 3H4, Canada}
\altaffiltext{5}{Dunlap Institute for Astronomy and Astrophysics, University of Toronto, Toronto ON, M5S 3H4, Canada}
\altaffiltext{6}{Canadian Institute for Theoretical Astrophysics, Toronto, ON, M5S 3H4, Canada}
\altaffiltext{7}{Special Astrophysical Observatory, Nizhnij, Arkhyz, Karachai-Cherkessia 369167, Russia}

\begin{abstract}
We obtained follow-up HST observations of the seven low surface brightness galaxies discovered with the Dragonfly Telephoto Array in the field of the massive spiral galaxy M101. 
Out of the seven galaxies, only three were resolved into stars and are potentially associated with the M101 group at $D=7\text{ Mpc}$.
Based on HST ACS photometry in the broad F606W and F814W filters, we use a maximum likelihood algorithm to locate the Tip of the Red Giant Branch (TRGB) in galaxy color-magnitude diagrams.  Distances are $6.38^{+0.35}_{-0.35}, 6.87^{+0.21}_{-0.30}$ and $6.52^{+0.25}_{-0.27} \text{ Mpc}$ and we confirm that they are members of the M101 group.
Combining the three confirmed low luminosity satellites with previous results for brighter group members, we find the M101 galaxy group to be a sparsely populated galaxy group consisting of seven group members, down to $M_V = -9.2 \text{ mag}$.
We compare the M101 cumulative luminosity function to that of the Milky Way and M31. We find that they are remarkably similar; In fact, the cumulative luminosity function of the M101 group gets even flatter for fainter magnitudes, and we show that the M101 group might exhibit the two known small-scale flaws in the $\Lambda\textrm{CDM}$ model, namely `the missing satellite' problem and the `too big to fail' problem.  Kinematic measurements of M101$'$s satellite galaxies are required to determine whether the `too big to fail'  problem does in fact exist in the M101 group. 

\end{abstract}

\keywords{Galaxy}

\section{Introduction}\label{intro}

Dwarf galaxies are the most abundant type of galaxy in the observed Universe (\citealt{1997AJ....113..185M}). They serve as unique cosmological probes and provide insight into many aspects of the formation and evolution of the Universe on small scales. 
%While the number of detected dwarf galaxies residing within the Local Group has increased significantly in recent years (e.g. \citealt{2012AJ....144....4M}; \citealt{2015ApJ...807...50B}; \citealt{2016MNRAS.459.2370T}), they are still very difficult to detect outside the Local Group due to their low surface brightness, low mass and small size. 
In a $\Lambda\textrm{CDM}$ concordance cosmology, the assembly history of galaxies is largely driven by the hierarchical growth of the underlying dark matter structures and by the merger and accretion of smaller halos (\citealt{1978MNRAS.183..341W}; \citealt{1985ApJ...292..371D}; \citealt{2008ApJ...683..597S} and references therein). 
Comparisons of observed dwarf galaxy abundances to predictions of $\Lambda\textrm{CDM}$ for the Milky Way, have led to two possible small-scale flaws in the model. The first, dubbed `the missing satellite' problem, refers to the apparent discrepancy between the predicted number of dark matter satellites in simulations and the observed number of dwarf galaxies around the Milky Way (\citealt{1993MNRAS.264..201K}; \citealt{1999ApJ...522...82K}; \citealt{1999ApJ...524L..19M}). A more concrete problem, namely the `too big to fail' problem, is concerned with dwarf galaxies at higher mass and luminosity. Dark matter only simulations predict too many massive dense sub-halos compared with satellites around the Milky Way (\citealt{2011MNRAS.415L..40B}; \citealt{2012MNRAS.422.1203B}).

The `missing satellite' and `too big to fail' problems are not unique to the Milky Way. Recent works have identified the `too big to fail' problem for Andromeda (M31) satellite galaxies (\citealt{2014MNRAS.440.3511T}), within the Local Group, not only within 300 kpc where environmental physics may be able to resolve the disagreement (\citealt{2014MNRAS.439.1015K}; \citealt{2014MNRAS.444..222G}) and also for field galaxies (\citealt{2015A&A...574A.113P}). 

%As of today, the underlying studies for these research areas are entirely based on data from the Local Group. Of course, it is interesting to consider whether the Local Group itself may be anomalous. Unfortunately, looking beyond the Local Group to discover whether the `too big to fail' and `missing satellite' problems are truly general is not easy.
As of today, the vast majority of the underlying studies for the `missing satellite' and the `too big to fail' problems are based on data from the Local Group. Of course, it is interesting to consider whether the Local Group itself may be anomalous. Unfortunately, looking beyond the Local Group to discover whether the `too big to fail' and `missing satellite' problems are truly general is not easy.
Star count surveys are only efficient at identifying faint dwarfs in the local universe ($\leq\textrm{2-3 Mpc} $; \citealt{2008ApJ...686..279K},\citealt{2009AJ....137..450W}), as the luminosity of stars falls with the square of the distance. 
Beyond $5 \text{ Mpc}$, galaxies are more easily detected in integrated light rather than star counts, as integrated surface brightness is independent of distance.  
%However, a key problem is that most dwarf galaxies in the Local Group have extremely low surface brightness and thus are difficult to observe, let alone dwarfs outside the Local Group. 
However, a key problem is that most dwarf galaxies within the Local Group have extremely low surface brightnesses, and are thus difficult to observe; and the problem is only intensified for dwarfs outside of the Local Group.
%Only a few dwarf galaxies have been identified outside of the Local Group by their low surface brightness emission (e.g. \citealt{2014arXiv1401.2719K}; \citealt{2014ApJ...787L..37M}).
%Even using telescopes that are highly optimized for low surface brightness imaging, one cannot easily observe low surface brightness below $\mu_B \sim 29 \text{ mag arcsec}^{-2}$ due to systematic errors in flat fielding and the complex wide-angle point spread functions of stars (\citealt{2009PASP..121.1267S}).
Even using telescopes that are highly optimized for low surface brightness imaging, one cannot easily push below $\mu_B \sim 29 \text{ mag arcsec}^{-2}$ due to systematic errors in flat fielding and the complex wide-angle point spread functions of stars (\citealt{2009PASP..121.1267S}; but see \citealt{2013AJ....146..126C}, \citealt{2014arXiv1401.2719K}, \citealt{2014ApJ...795L..35C} and \citealt{2016ApJ...823...19C} for examples of detected low surface brightness dwarfs associated with other galaxy groups).
%If the known Milky Way satellites were located at 5 Mpc, beyond the reach of star count studies, their median central surface brightness would be $\mu_{V} \sim 26.1 \textrm{ mag arcsec}^{-2}$, too faint to detect in most surveys of unresolved integrated light. 

To address this problem, a new robotic refracting telescope was developed and built, called Dragonfly, that is optimized to explore the universe at surface brightness levels below $\mu_{B} = 30 \textrm{ mag arcsec}^{-2}$ (\citealt{2014PASP..126...55A}). When the observations this research is based on were taking place, the telescope was comprised of an array of eight 400 mm $f/2.8$ Canon IS II telephoto lenses which all image only slightly offset fields of view, forming effectively a 0.4 m aperture $f/1.0$ refractor. The telescope takes simultaneous $g$ and $r$ images over an instantaneous $2.6^{\circ} \times 1.9^{\circ}$ field of view. 
The first target of the Dragonfly array was M101, the nearest massive spiral galaxy beyond M31 (\citealt{2014ApJ...782L..24V}). \citealt{2014ApJ...787L..37M} identified seven large, low surface brightness (LSB) objects in the field of M101. 
All seven candidates lie within the M101 projected virial radius and assuming the distance to M101, their physical properties (color, size, luminosity and morphology) are consistent with Local Group dwarf satellite galaxies, as shown in \citealt{2014ApJ...787L..37M}. 

In this paper, we describe follow-up observations with the \textsl{Hubble Space Telescope} (HST) Advanced Camera for Surveys (ACS) images in the broad F606W and F814W for all seven dwarfs, to generate Color-Magnitude Diagrams (CMDs) from which to establish membership based on the Tip of the Red Giant Branch (TRGB) method.
The primary goal of the observation is to measure distances to the seven dwarfs and to test whether they are members of the M101 group. 
%Using the determination of M101 group companionship and considering the extremely low limiting surface brightness in the reduced images, a mindful discussion is given regarding a possible `missing satellite' problem and `too big to fail' problem of the M101 group. 
Using the determination of M101 group companionship and considering the extremely low limiting surface brightness in the reduced images, our results shed some light on the possible `missing satellite' and `too big to fail' problems in the context of the M101 group.

\section{Observations and Data Reduction}\label{data}

\begin{figure*}[t]
\centering
  \includegraphics[width=160mm]{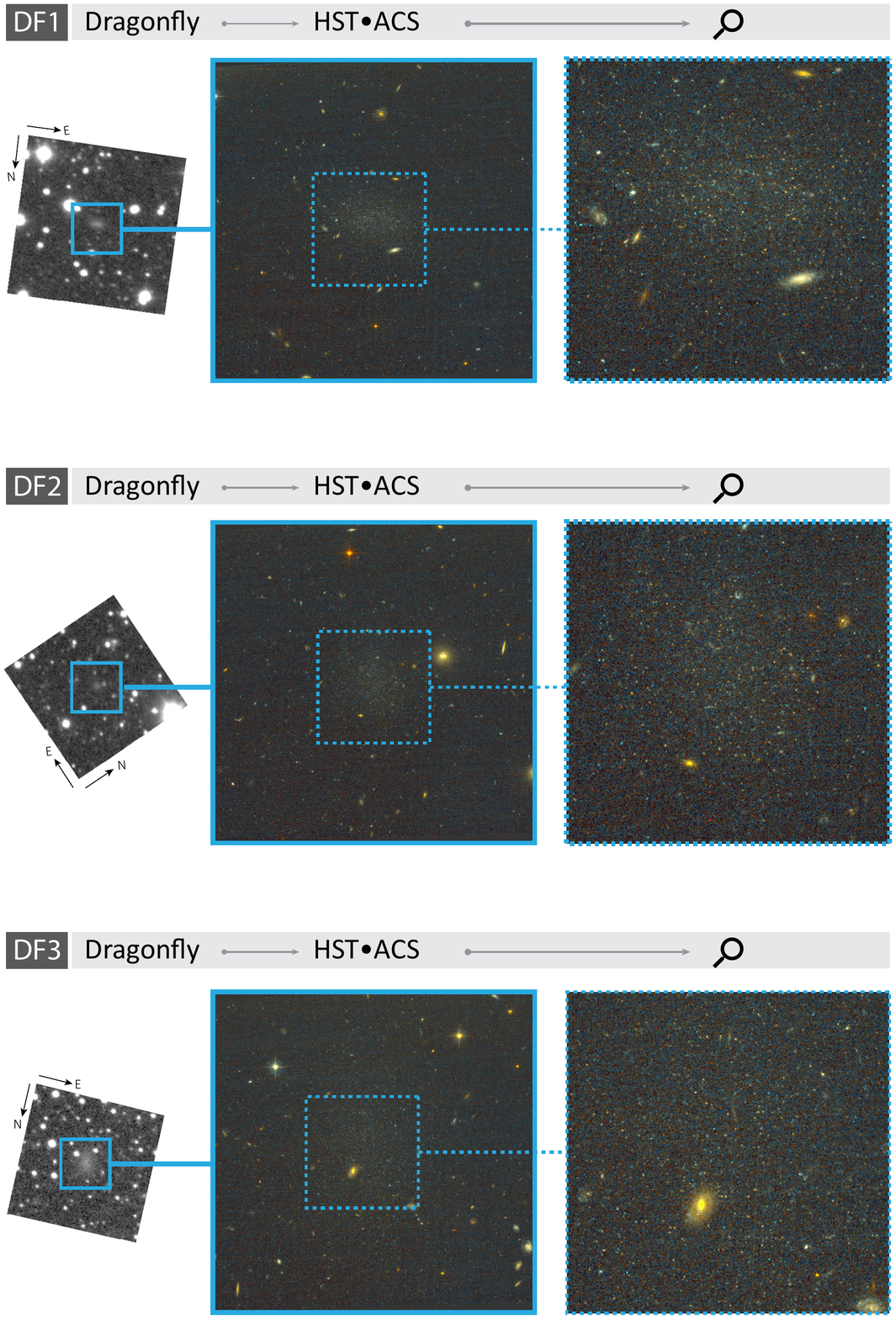}
%  \caption{Color images of 7 dwarfs (left to right) observed with HST ACS WFC camera.}
  \caption{The three Dragonfly-discovered galaxies in the M101 field, observed with Dragonfly $g$-band (left) and HST/ACS (center). The right image for each object shows that the galaxies are resolved into a wealth of stars with the ACS resolution.}
  \label{fig:hst_col}
\end{figure*}

\subsection{HST Observations}\label{HST}
Establishing group membership requires accurate distances to each of the seven satellite candidates discovered in \citealt{2014ApJ...787L..37M}. The Dragonfly images and other available CFHT imaging do not resolve the galaxies into stars, which makes it difficult to measure their distances. 
If the galaxies are at the distance of M101 or closer (at $\sim 7 \text{ Mpc}$), we expect the brighter members of galaxy stellar populations, including red giant stars, to be resolved into individual stars in single orbit ACS imaging. Indeed, only 0.5 orbits per filter were needed to obtain a distance estimate using the well-known TRGB method (e.g., \citealt{2008ApJ...688.1009M}; \citealt{2009ApJS..183...67D}; \citealt{2011ApJS..195...18R}). The TRGB has $M_{I,\text{TRGB}} \approx -4.0$, and if the galaxies are at the distance of M101, the observed TRGB is at $I \approx 25 \text{ mag}$. With the ACS on HST, we can reach $I = 26.8 \text{ mag}$  with $5\sigma$, almost 2 magnitudes below the tip, assuring secure determination of the TRGB. 

The seven satellite candidates were observed with the HST ACS, Wide Field Channel (WFC)\footnote{HST/ACS cycle 22, proposal ID: 13682.}. Four exposures were made in a single orbit per galaxy: F606W (broad-band V), F814 (broad-band I) filters at the same position as well as F606W and F814W at a dithered position, $\sim 3''$ away. The total exposure time per galaxy for each filter was 1150 seconds.
The ACS/WFC data was reduced with the default HST reduction pipeline. 

Remarkably, only three out of the seven LSB galaxies, DF1, DF2 and DF3, were resolved into stars in the HST ACS images. The rest of the analyses and results presented in the paper refer only to the three resolved galaxies. 
%Merritt et al. 2016 discuss the other four unresolved objects. 
Merritt et al. 2016 discuss the other four unresolved objects and consider them to be most likely associated with a background group at $\sim 26 \text{ Mpc}$, containing NGC 5485 and NGC 5473.
Figure \ref{fig:hst_col} shows a panel of the Dragonfly $g-$band and HST color images for each of the three resolved dwarf galaxies. For each HST image, we also show the rich collection of stars resolved with the ACS resolution. 

%\begin{figure*}[t]
%\centering
%  \includegraphics[width=180mm]{figures/cfht_fig.png}
%  \caption{Color images of DF1, DF2 and DF3 observed with CFHT.}
%  \label{fig:cfht_col}
%\end{figure*}

\subsection{Stellar Photometry}\label{photometry}
To measure stellar photometry, we use the publicly available ACS module of DOLPHOT, a modified version of HSTphot (\citealt{2000PASP..112.1383D}). The photometry was carried out on each of the bias subtracted, flat-fielded, CTE-corrected *.flc images produced by the STScI ACS pipeline. The ACS module of DOLPHOT provides a set of pre-processing utilities including bad column and hot pixel masking, cosmic-ray rejection, sky determination and images alignment. 
These pre-processing steps are followed by Point Spread Function (PSF) fitting photometry with model TinyTim PSFs, generated by DOLPHOT.
%We configured DOLPHOT with parameters similar to those used in the ACS Nearby Galaxy Survey Treasury program (\citealt{2009ApJS..183...67D}) and the GHOSTS survey (\citealt{2011ApJS..195...18R}). In particular, DOLPHOT photometry quality was maximized when we used the parameters: FitSky=3, RAper=10 (pixels) and Force1=1. More about the parameters selection is described in DOLPHOT manuals. 
DOLPHOT parameters were set as recommended in the manual for ACS/WFC data. Three of the parameters which had the strongest influence on the resulted photometry were the sky fitting parameter (FitSky), the aperture radius (RAper) and the Force1 parameter which forces all sources detected to be fitted as stars. These parameters were extensively studied by the ACS Nearby Galaxy Survey Treasury program (\citealt{2009ApJS..183...67D}) for crowded regions and we adopted their values (FitSky=3, RAper=10 pixels and Force1=1. See the DOLPHOT manual and \citealt{2009ApJS..183...67D} for more details).

The final output of DOLPHOT lists the position of each detected source as well as object type (point source, extended, elongated or intermediate), measured Vega magnitudes and their corresponding errors, count rate, rate error, background, $\chi^2$ for the fit of the PSF, sharpness, roundness, crowding, signal-to-noise value and quality flag. High quality catalogs for the three resolved dwarfs were produced with stellar photometry that have been culled to remove highly uncertain photometry. The catalogs have been filtered to those objects that are considered point sources (object type=1) with $\text{quality-flag}\leq2$, sharpness: $(\text{sharp}_{\text{F606W}}+\text{sharp}_{\text{F814W}})\leq0.075$, crowding: $(\text{crowd}_{\text{F606W}}+\text{crowd}_{\text{F814W}})\leq0.1$ and S/N greater than 4.0 in both filters.

\begin{figure*}
\centering
  \includegraphics[width=150mm]{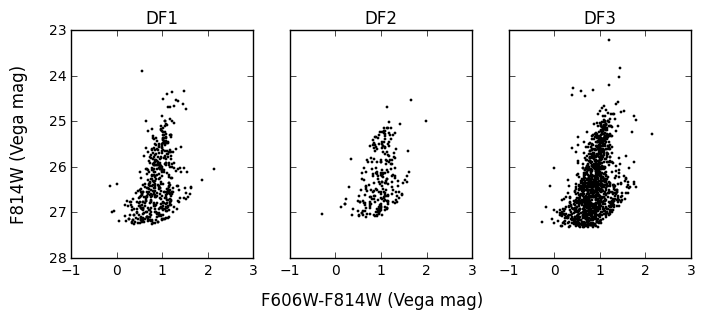}
  \caption{CMDs within one effective radius for the three well-resolved candidate M101 satellite galaxies imaged with HST ACS and photometered with DOLPHOT. The CMDs are well-populated and the red giant branch is evident in all three.}
  \label{fig:cmd1_3}
\end{figure*}

%CMDs for the three well resolved satellite candidates and for the four non-resolved candidates are shown in Figures \ref{fig:cmd1_3} and \ref{fig:cmd4_7} respectively. The plotted photometry is drawn from the high-quality catalogs. The detected sources in the four non-resolved target's CMDS include a non-negligible number of detections of noise and nearby galaxies and therefore not reliable for further TRGB analysis.
CMDs for the three well resolved satellite candidates are shown in Figure \ref{fig:cmd1_3}. The plotted photometry is drawn from the high-quality catalogs.

%The CMDs for the resolved dwarfs 
%Error bars denote $1\sigma$ uncertainties in magnitude and color as estimated from the artificial star tests. In the 3 resolved LSBs the Red Giant Branch (RGB) is visible, however in the 
%The CMD is not featured since the image was not resolved. 

%\begin{figure*}
%\centering
%  \includegraphics[width=150mm]{figures/cmd_df4_7.png}
%  \caption{CMDs within one effective radius for the four unresolved targets imaged with HST ACS and photometered with DOLPHOT. Despite the high-quality photometry culls applied to the stars catalogs, many detections are either %noise or parts of nearby galaxies. Lower limits on their distances are placed in Merritt et al. 2016 (in prep).}
%  \label{fig:cmd4_7}
%\end{figure*}

\section{Distance measurements}\label{results}
\subsection{The TRGB Method}\label{TRGB}
The method of deriving the distance from the Tip of the Red Giant Branch (TRGB) magnitude is arguably the most valuable distance indicator for galaxies within ~10 Mpc. It offers a clear and well understood physical basis (\citealt{1997MNRAS.289..406S};  \citealt{1997eds..proc..239M} and \citealt{2002PASP..114..375S}): in the course of evolution of low- to intermediate-mass stars, they turn into red giants by gradually increasing their luminosity. Once the helium in the core of the star reaches a temperature of roughly $10^8\text{ K}$, the entire core begins helium fusion nearly simultaneously in a short intense helium flash. It then contracts and leaves the red giant branch.

Observationally, this physical phenomenon causes a sharp cutoff of the bright end of the red giant branch luminosity function (LF), approximately located at $M_{I,\text{TRGB}} \approx -4.0$. The first to notice the similar luminosity of the brightest red giants of galaxies in the Local Group was Baade (\citeyear{1944ApJ...100..137B}). Later on, another study by Sandage (\citeyear{1971ApJ...166...13S}) confirmed these observations for a large number of Local Group galaxies. He found that the absolute magnitude of the brightest red giant stars in the underlying sheets of M31, M33 and IC 1613 peaks at the same magnitude of $M_V \approx -3.0 \pm 0.2 \textrm{ mag}$. 

We use the latest development of the Maximum Likelihood method for characterizing the magnitude of the TRGB presented by \citealt{2006AJ....132.2729M}. It offers several improvements over other methods that were used in the past (\citealt{1993ApJ...417..553L}, \citealt{1996ApJ...461..713S} and \citealt{2002AJ....124..213M}). In this method, no binning or smoothing is required and it accounts for systematic photometric errors, which reduces the total error significantly. 
\citealt{2006AJ....132.2729M} presented two improvements in comparison to the previous offered Maximum Likelihood method (\citealt{2002AJ....124..213M}). The first is the parametrization of the red giant branch LF. The fitting function to the LF is assumed to be a simple power-law with a cutoff in the TRGB region plus a power-law of a second slope for a stellar population brighter then the TRGB
\begin{equation}
\psi = 
  \left\{
	\begin{array}{c}
	10^{a\left(m-m_{\text{TRGB}}\right)+b},\quad m-m_{\text{TRGB}}\geq0\\
	10^{c\left(m-m_{\text{TRGB}}\right)},\quad m-m_{\text{TRGB}}<0
	\end{array}
  \right.
\end{equation}

The second improvement presented by \citealt{2006AJ....132.2729M} is the use of a photometric error function defined from artificial star tests. The procedure of artificial star tests involves the generation of a very large library of artificial stars that cover the necessary range of magnitudes and colors on the CMD such that the distribution of recovered photometry is adequately sampled. The photometry program, DOLPHOT, analyzes the generated artificial stars with the same routines and parameter selections as for observed stars. Photometric error estimations are determined by comparison of predefined input magnitudes with photometered magnitudes. The smoothed LF is then given by 
\begin{equation}
\varphi(m) = \int \psi(m')\rho(m')e(m\mid m')dm'
 \end{equation}
where $\rho(m)$ is the completeness function and $e(m\mid m')$ is the error distribution. For the latter a Gaussian distribution is assumed and take into account the bias in the photometric error function.

\subsection{Results}\label{results}
As mentioned in subsection \ref{HST}, only three out of the seven satellite candidates were well resolved into stars and thus allowed us to use the TRGB method as a distance indicator for these galaxies. 

In order to determine the apparent magnitude of the tip in the I-band, $m_I$ ,we use the TRGBTOOL program, written and provided by D.Makarov and ran it on the DOLPHOT photometry. The algorithm implements the Maximum Likelihood method described in \citealt{2006AJ....132.2729M} and was optimized by introducing reliable photometric errors and a completeness characterization determined with the artificial star tests. 
The absolute magnitude of the TRGB in the I-band was estimated using the calibration of the TRGB from \citealt{2007ApJ...661..815R}. They define a zero-point calibration of the TRGB, accurate to $1\%$ statistical uncertainty, as a function of the stellar population color to account for variation due to metallicity and age. The zero-point is provided in the HST flight system for $F606W$ and $F814W$ for ACS. Their result is given by 
\begin{equation}
M^{\text{ACS}}_{F814W} = -4.06 + 0.20[(F606W - F814W)-1.23]
 \end{equation}
The estimation of the TRGB can be made directly within the flight system magnitudes as measured by DOLPHOT. 

\begin{figure*}[t]
\centering
  \includegraphics[width=115mm]{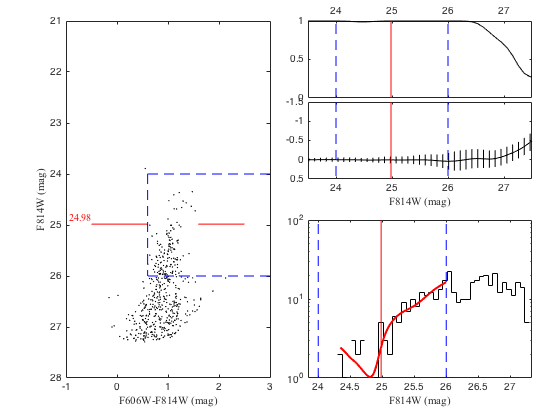}
  \caption{Color Magnitude Diagram (\textit{left panel}) and TRGB calculation results (\textit{right panel}) for DF1. The top right panel shows the completeness, photometric errors and dispersion in errors vs. the HST ACS F814W band magnitude. The bottom right panel shows a histogram of the F814W LF (black) and the resulting model LF convolved with photometric errors and incompleteness (red). The vertical red line denote the position of the TRGB.}
  \label{fig:cmd_df1}
\end{figure*}

\begin{figure*}[t]
\centering
  \includegraphics[width=115mm]{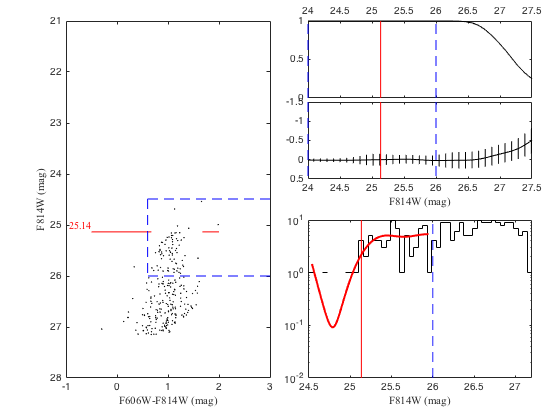}
  \caption{Color Magnitude Diagram (\textit{left panel}) and TRGB calculation results (\textit{right panel}) for DF2. The top right panel shows the completeness, photometric errors and dispersion in errors vs. the HST ACS F814W band magnitude. The bottom right panel shows a histogram of the F814W LF (black) and the resulting model LF convolved with photometric errors and incompleteness (red). The vertical red line denote the position of the TRGB.}
  \label{fig:cmd_df2}
\end{figure*}

\begin{figure*}[t]
\centering
  \includegraphics[width=115mm]{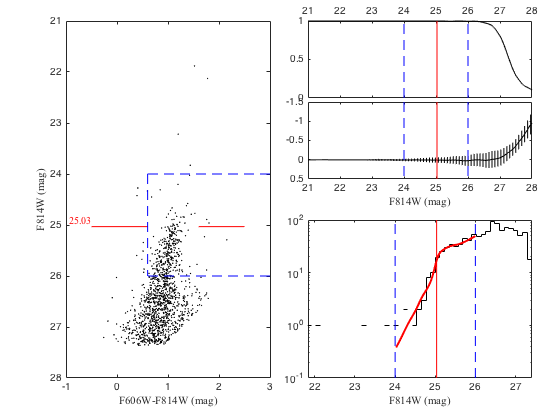}
  \caption{Color Magnitude Diagram (\textit{left panel}) and TRGB calculation results (\textit{right panel}) for DF3. The top right panel shows the completeness, photometric errors and dispersion in errors vs. the HST ACS F814W band magnitude. The bottom right panel shows a histogram of the F814W LF (black) and the resulting model LF convolved with photometric errors and incompleteness (red). The vertical red line denote the position of the TRGB.}
  \label{fig:cmd_df3}
\end{figure*}

The CMDs and TRGB calculation results for DF1, DF2 and DF3 are presented in Figures \ref{fig:cmd_df1}, \ref{fig:cmd_df2} and \ref{fig:cmd_df3}. The plotted photometry (left panel) is drawn from the high quality catalogs produced after applying the photometry cuts, as described in subsection \ref{photometry}. The photometric errors, dispersion in errors and completeness (upper right panel) were estimated using the artificial star tests. 
The lower right panel shows the LF and the resulting model LF convoluted with the photometric errors and incompleteness.

The galaxies DF1, DF2 and DF3 are located at distances of $6.37^{+0.35}_{-0.35}\text{ Mpc}$, $6.87^{+0.21}_{-0.30}\text{ Mpc}$ and $6.52^{+0.25}_{-0.27}\text{ Mpc}$ respectively. More than 80 distance measurements for M101 can be found on the NASA Extragalactic Database (NED) where the mean distance is $7.0\text{ Mpc}$ and its median distance value is $6.9\text{ Mpc}$.\citealt{2015AstL...41..239T} present stellar photometry for several fields around M101 and use the TRGB method to determined the distance to these fields to have a mean value of $6.79^{+0.11}_{-0.11}\text{ Mpc}$.
Based on the high accuracy TRGB distance determination to the resolved LSB dwarfs, we conclude that they are members of the M101 group and thus confirm a total of three new M101 companion dwarf satellites of the seven initially discovered Dragonfly objects.
Even though the galaxies are all within the projected virial radius and their TRGB distances are all consistent within $\sim 1 \sigma$ with the M101 distance, we cannot exclude that the 3D positions place them outside it.

\section{The M101 Group}
\subsection{Known Members of the M101 Group}

%In order to talk about the `missing satellite' problem in the M101 group, we need to be sure we are accurately determining the members in this galaxy group. 
Over the past $\sim 40$ years, many studies have focused on the M101 group and among other objectives, tried to determine its possible companions (\citealt{1978A&A....64..359A}; \citealt{1982ApJ...257..423H}; \citealt{1983ApJS...52...61G}; \citealt{1988ang..book.....T}; \citealt{1993A&AS..100...47G}; \citealt{1994BSAO...38....5K}; \citealt{2011MNRAS.412.2498M}; \citealt{2014AJ....148...50K}; \citealt{2014ApJ...787L..37M}; \citealt{2015AstBu..70..379K}; \citealt{2015AstL...41..239T}).
%\citealt{1994BSAO...38....5K} measured photometric distances with large errors, ending up in overestimating the number of companions of M101. Later studies by Makarov and Karachentsev (\citeyear{2011MNRAS.412.2498M}) and Karachentsev and Kudrya (\citeyear{2014AJ....148...50K}) determined the group composition based on similarity of the radial velocities of the galaxies, a questionable method for determining group membership. Thus, accurate measurements of the distances to satellite candidates were needed to identify the true companions of M101.
\citealt{1994BSAO...38....5K} measured photometric distances with large errors, ending up in overestimating the number of companions of M101. Later studies by Karachentsev and Kudrya (\citeyear{2014AJ....148...50K}) determined the group composition by combining all galaxies with a common ``Main Disturber'' (MD) into an association, based on the magnitude of the tidal forces and assuming that all the companions of the MD are the same distance from the observer as the MD itself. The latter assumption can lead to large uncertainties in group membership if the companions have very different distances from the observer.
Thus, accurate measurements of the distances to satellite candidates were needed to identify the true companions of M101.
Such an analysis was carried by \citealt{2015AstL...41..239T} which used the TRGB method to determine the distances to dwarf galaxies in the surroundings of M101, M51 and M63. Their distance measurements to galaxies in the neighborhood of M101 resulted in four dwarf companions: NGC 5474, NGC 5477, UGC 8837 and UGC 9405. We consider Tikhonov's result to be the most credible due to the high accuracy of the TRGB method. 
%and thereby the conclusion of physical membership of a measured galaxy in a particular group.

\begin{figure*}
	\centering
		\includegraphics[width=160mm]{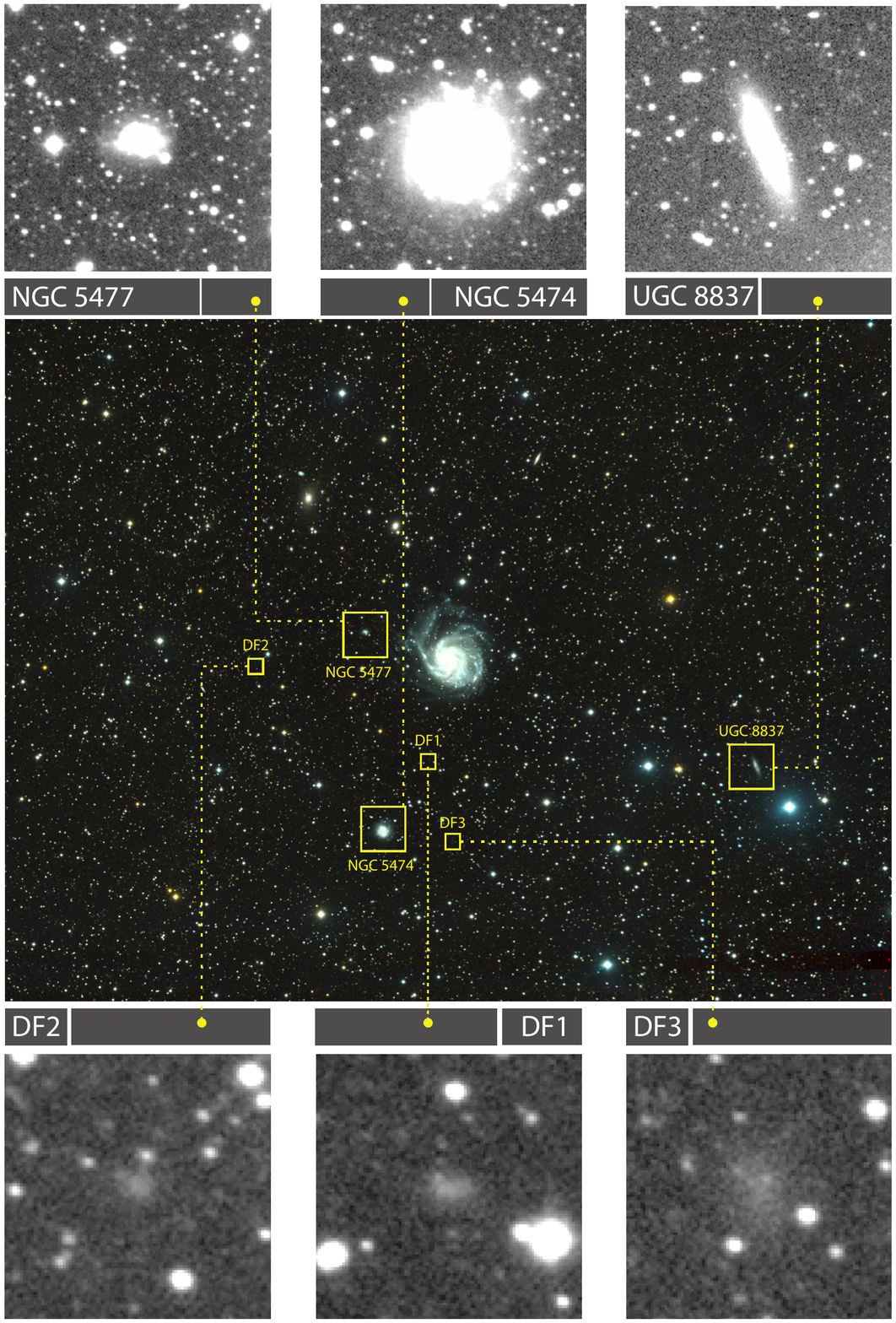}
%		\includegraphics[width=180mm]{figures/fov_lines.png}
%		\caption{The full  $3.3^{\circ} \times 2.8^{\circ}$ Dragonfly field of view color image, centered on M101. The zoomed cutouts highlight the position of each of the seven low surface brightness objects, discovered in \citealt{2014ApJ...787L..37M}.}
		\caption{The full  $3.3^{\circ} \times 2.8^{\circ}$ Dragonfly field of view color image, centered on M101. The upper and lower panels show the position and enlargement of the $g-$band Dragonfly images of six out of seven M101 satellites. DF1, DF2 and DF3 (bottom) were first discovered in \citealt{2014ApJ...787L..37M} and determined to be members of the M101 group in this study. Three out of the four already known members of the group, NGC 5474, NGC 5477 and UGC 8837 (\citealt{2015AstL...41..239T}) fall within the Dragonfly field of view (top). The fourth known member of M101, UGC 9405, doesn't lie within its projected virial radius.}
 		\label{fig:fov_DF}
\end{figure*}

%The Dragonfly image is shown is Figure \ref{fig:fov_DF} (taken from \citealt{2014ApJ...787L..37M}).
The full field of view Dragonfly image, centered on M101, is shown is Figure \ref{fig:fov_DF}. The zoomed cutouts show the current six out of seven satellites of the M101 group: DF1, DF2 and DF3 (bottom), presented as group companions in this study, as well as three out of four already known companions of the M101 group, NGC 5474, NGC 5477 and UGC 8837 (top), presented in \citealt{2015AstL...41..239T}. The fourth known member, UGC 9405, falls outside the Dragonfly field of view. 
In Table \ref{table:results} we summarize the results for all the satellite galaxies of M101: the apparent magnitude of the TRGB in F814W, the absolute magnitude of the TRGB in F814W, the TRGB color index, the resulting distance modulus and finally the inferred distance in $\text{Mpc}$.   
%from the TRGB analysis carried out in this work for DF1, DF2 and DF3
%\setlength{\tabcolsep}{1pt}

%\noindent\makebox[\textwidth]{}
\begin{deluxetable*}{ccccccccccc}\label{table:results}
\tabletypesize{\scriptsize}
\tablewidth{0.99\textwidth}
\centering
\tablecolumns{10}
\tablecaption{Photometry, TRGB and distance Measurements for M101 group satellites\label{table:results}}
\tablehead{   % column headings
  \colhead{Target} &
  \colhead{$\alpha$} &
  \colhead{$\delta$} &
%  \colhead{$\mu_{0,g}$\,\tablenotemark{a}} &
  \colhead{$\mu_{e,g}$\,\tablenotemark{b}} &
   \colhead{$m_{I,\text{TRGB}}$\,\tablenotemark{c}} &
  \colhead{Mean\,\tablenotemark{d}} &
  \colhead{$M_{I,\text{TRGB}}$\,\tablenotemark{e}} &
  \colhead{$(m-M)_0$\,\tablenotemark{f}} &
  \colhead{$(m-M)_0,\text{ext}$\,\tablenotemark{g}} &
  \colhead{$D$\,\tablenotemark{h}}\\
  \colhead{Name}&{(J2000)}&{(J2000)}& & && \colhead{Color} && &(Mpc)
}

\startdata
M101-DF1&14 03 45.0&+53 56 40&$26.6\pm0.1$&$24.98^{0.12}_{0.12}$&$1.09^{0.05}_{0.06}$&$-4.08^{0.01}_{0.01}$&29.06&$29.02^{0.12}_{0.12}$&$6.37^{0.35}_{0.35}$\\  \\

M101-DF2&14 08 37.5&+54 19 31&$26.9\pm0.2$&$25.13^{0.06}_{0.09}$&$1.17^{0.06}_{0.08}$&$-4.07^{0.01}_{0.01}$&29.20&$29.19^{0.06}_{0.09}$&$6.87^{0.21}_{0.30}$\\  \\

M101-DF3&14 03 05.7&+53 36 56&$27.4\pm0.2$&$25.03^{0.08}_{0.09}$&$1.10^{0.02}_{0.08}$&$-4.086^{0.003}_{0.016}$&29.12&$29.07^{0.08}_{0.09}$&$6.52^{0.25}_{0.27}$\\  \\

NGC 5474&14 05 01.6&+53 39 44&$$&$25.13$&$1.33$&$-4.04$&$$&$29.17^{0.13}_{0.13}$&$6.82$\\  \\

NGC 5477&14 05 33.3&+54 27 40&$$&$25.13$&$1.43$&$-4.02$&$$&$29.15^{0.13}_{0.13}$&$6.77$\\  \\

UGC 8837&13 54 45.7&+53 54 03&$$&$25.15$&$1.28$&$-4.05$&$$&$29.20^{0.15}_{0.15}$&$6.93$\\  \\

UGC 9405&14 35 24.1&+57 15 21&$$&$25.0$&$1.53$&$-4.0$&$$&$29.00^{0.13}_{0.13}$&$6.30$\\  \\

\enddata
\tablecomments{The TRGB magnitudes and distances for the Dragonfly galaxies, DF1, DF2 and DF3, were calculated in this study. 
The values for the rest of M101 satellites are from \citealt{2015AstL...41..239T}.}
\tablenotetext{a}{Integrated apparent magnitude, \citealt{2014ApJ...787L..37M}.}
\tablenotetext{b}{Central surface brightness, in $\text{mag arcsec}^{-2}$, \citealt{2014ApJ...787L..37M}.}
\tablenotetext{c}{TRGB apparent magnitude, in F814W.}
\tablenotetext{d}{TRGB mean color index.}
\tablenotetext{e}{TRGB absolute magnitude, in F814W.}
\tablenotetext{f}{Distance modulus.}
\tablenotetext{g}{Distance modulus corrected for extinction.}
\tablenotetext{h}{Distance, in Mpc.}

\end{deluxetable*}

\subsection{The Observed Cumulative Luminosity Function of the M101 Group}

\begin{figure}[t]
  \includegraphics[width=\linewidth]{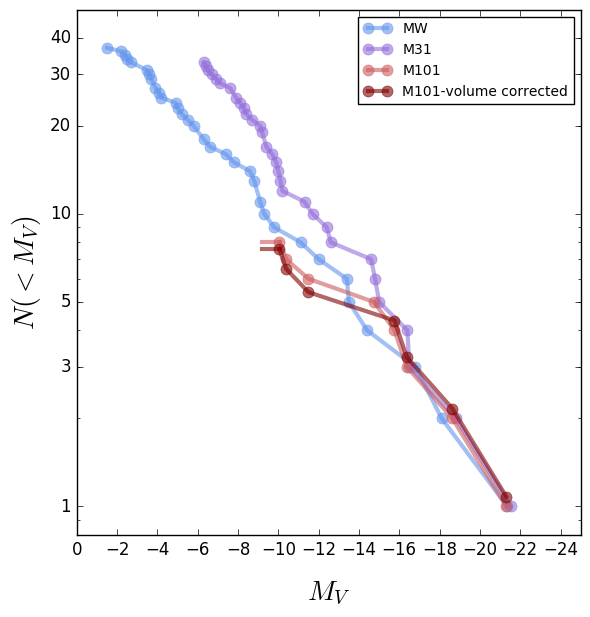}
  \caption{Observed cumulative luminosity function of the M101 group (red), including the LSBs presented here. The dark red curve represents the cumulative luminosity function of the six (out of a total of seven) M101 satellites that are inside the projected virial radius, multiplied by 1.08 to account for the virial volume outside the dragonfly field. 
We compare this to the observed CLFs of the Milky Way in blue (\citealt{2012AJ....144....4M};\citealt{2015ApJ...807...50B}; \citealt{2016MNRAS.459.2370T})) and M31 in purple (\citealt{2012AJ....144....4M}). The end point of the horizontal red line denotes the limiting absolute magnitude of the field around M101, using the Dragonfly array.}
  \label{fig:clf}
\end{figure}

The discovery of new Dragonfly low surface brightness ($-11.5<M_{V}<-10$) dwarfs in \citealt{2014ApJ...787L..37M} and the determination of their companionship to the M101 group in this work, can be used to produce the observed cumulative luminosity function, down to very low magnitudes. As mentioned in section \ref{results},  based on the high accuracy TRGB distance determination to the resolved dwarfs (DF1-DF3), we confirm a total of three new M101 companion dwarf satellites of the seven initially discovered Dragonfly objects. Therefore, to the best of our knowledge, the M101 group consists of seven companions with $-19<M_{V}<-10$. 
%Even though the projected virial radius of M101, $260 \text{ kpc}$, falls within the Dragonfly field of view, $392 \times 310 \text{ kpc}$, we cannot rule out other group members outside this field (as UGC 9405). It will be interesting to cover a wider field around M101 with similar and even lower surface brightnesses. 
Since the Dragonfly field of view, $392 \times 310 \text{ kpc}$, doesn't fully cover the projected virial radius of M101, we cannot rule out other group members outside this field (as UGC 9405).
In order to account for this volume incompleteness, we assume an NFW (\citealt{1997ApJ...490..493N}) distribution for the satellites and calculate a correction factor by which the cumulative number of satellites inside the virial volume has to be multiplied, in order to get a coverage complete representation of the luminosity function. Carrying out this calculation the correction factor is 1.08, which results in a total number of 6.48 satellites in the group, compared to 6 observed inside the projected virial radius.
It will be interesting to cover a wider field around M101 with similar and even lower surface brightnesses. 

In Figure \ref{fig:clf}, we show the cumulative luminosity function for the M101 group, along with the observed cumulative luminosity functions of the Milky Way and M31 for comparison. 
For completeness purposes, we show both the M101 cumulative luminosity function that includes the seven members of the group observed within the Dragonfly field of view, as well as the corrected cumulative luminosity function, taking into account the six satellites lie inside the virial volume of the group multiplied by 1.08 to account for the virial volume outside the Dragonfly field.
The three galaxies in the M101 group with the lowest magnitudes are the new dwarf companions investigated in this paper. The end point of the horizontal red line denotes the limiting absolute magnitude of the field around M101, using the Dragonfly array. The limiting absolute magnitude in the field around M101 was estimated conservatively by taking into account the measured apparent magnitude of the faintest detected object in that field using the Dragonfly array, DF7, with an integrated apparent magnitude of $m_g = 20.4 \text{ mag}$. 
%, taking into account an object with $r_e \sim 350 \text{ pc}$ ($10 \text{ arc seconds}$ in the distance of M101).

For $M_V \lesssim -15$ the observed cumulative luminosity functions of satellites around Milky-Way, M31 and M101 are remarkably similar. Fainter than $M_V \sim -15$ the cumulative luminosity function starts to flatten up to $M_V \approx -9$. The number of satellite galaxies for magnitudes fainter than $M_V \approx -10$ is significantly lower than this of Milky Way and M31 for the same magnitudes. 

\subsection{$V_{\text{max}}$ for the M101 Group Members}
%\subsection{The Cumulative $V_{\text{max}}$ Function of the M101 Group}

\begin{figure}[t]
  \includegraphics[width=\linewidth]{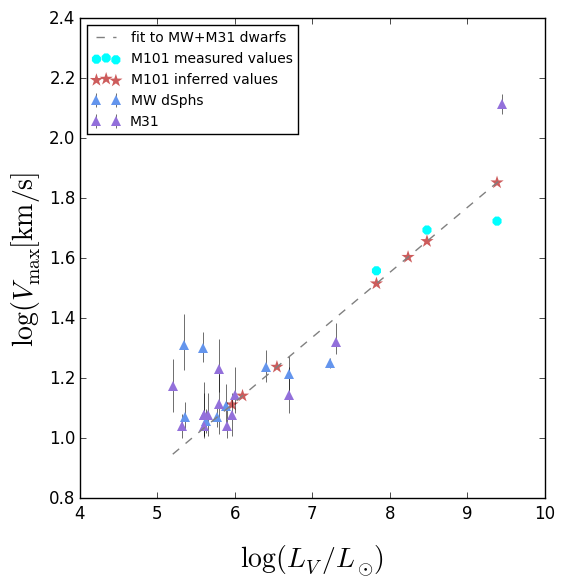}
  \caption{The luminosity-halo circular velocity relation of dwarf galaxies. The triangles are dSphs, blue from Milky Way (\citealt{2012MNRAS.422.1203B}), and purple from M31 (\citealt{2014MNRAS.440.3511T}). The dashed line is a fitted power law of the Milky Way and M31 dwarfs and the red stars are the inferred values for the M101 dwarfs, using the obtained relation. We also show measured values of the three out of the four brightest M101 already known satellites of M101 group as cyan circles (\citealt{1997ApJS..112..315H}; \citealt{2013MNRAS.433L..30L}; \citealt{2015MNRAS.451.3021D}; \citealt{2014A&A...561A.131S}; \citealt{2005ApJ...625..763L}).}
  \label{fig:lum_vmax_rel}
\end{figure}

By examining the luminosity-halo circular velocity relation of dwarf galaxies of Milky Way and M31, \citealt{2014MNRAS.440.3511T} showed that even though Milky Way and M31 dwarfs do not appear to be well described by abundance matching, nor its extrapolation, they do present monotonically increasing halo mass with increasing luminosity.
We take this finding one step further by simply fitting a power-law to the relation between the halo maximum circular velocity, $V_{\text{max}}$ and the luminosity,$L/L_{\odot}$, to the satellites of Milky Way and M31. 

We use the relation between $V_{\text{max}}$ and $L/L_{\odot}$ to infer the values of the circular velocities for the seven M101 companions.
In Figure \ref{fig:lum_vmax_rel} we show the maximum circular velocity-luminosity relation of observed Milky Way and M31 dwarfs (as blue and purple triangles, respectively) as well as the M101 dwarfs values (as red stars) inferred from the fitted power-law model (dashed line). 
We also show measured values of $V_{\text{max}}$ for three out of the four brightest, already known members of the M101 group as cyan circles: \citealt{1997ApJS..112..315H} inferred the maximum circular velocity of NGC 5474 from the width of its integrated HI emission profile, \citealt{2015MNRAS.451.3021D} and \citealt{2014A&A...561A.131S} measured the maximum circular velocity of UGC 8837 from rotation curves, and \citealt{2005ApJ...625..763L} obtained the maximum circular velocity from HI observations.
% (\citealt{1997ApJS..112..315H}; \citealt{2013MNRAS.433L..30L}; \citealt{2015MNRAS.451.3021D}; \citealt{2014A&A...561A.131S}; \citealt{2005ApJ...625..763L}). 
The measured values of the three brightest dwarfs of the M101 group lay within the confidence interval of the inferred values. This suggests that the power-law model we obtained is a reasonable approximation for the $V_{\text{max}}$ values of the M101 satellites. 
%It is also clear from Figure \ref{fig:lum_vmax_rel} that the satellite galaxies of M101 span a wide range of luminosities where most of the satellites have higher luminosities than Milky Way and M31 satellites. This suggests that the M101 group might have different characteristics shaping its formation and evolution.
We discuss further related results in \ref{tbtf}.

\section{Discussion}
\subsection{A `Missing Satellite' Problem for M101}
%Cosmological N-body simulations have been a very successful tool for testing predictions of the $\Lambda$CDM model of galaxy formation and evolution in last few decades (e.g. \citealt{1985ApJ...292..371D}; \citealt{1988ApJ...327..507F}; \citealt{1992ApJ...399..405W}; \citealt{1994ApJ...436..467G}; \citealt{1994ApJ...431..451C}; \citealt{1996ApJ...457L..51H}; \citealt{1998MNRAS.301...81G}; \citealt{2001MNRAS.321..372J}; \citealt{2004ApJ...606L..93W}; \citealt{2005MNRAS.364.1105S}; \citealt{2009MNRAS.398.1150B}; \citealt{2011ApJ...740..102K}). The formation and evolution of cosmic large scale structure has been studied thoroughly and with a remarkable agreement to a diversity of numerical simulations. 
Cosmological N-body simulations have been a very successful tool for testing predictions of the hierarchical galaxy formation model in a $\Lambda$CDM cosmology in the last few decades (e.g. \citealt{1985ApJ...292..371D}; \citealt{1988ApJ...327..507F}; \citealt{1992ApJ...399..405W}; \citealt{1994ApJ...436..467G}; \citealt{1994ApJ...431..451C}; \citealt{1996ApJ...457L..51H}; \citealt{1998MNRAS.301...81G}; \citealt{2001MNRAS.321..372J}; \citealt{2004ApJ...606L..93W}; \citealt{2005MNRAS.364.1105S}; \citealt{2009MNRAS.398.1150B}; \citealt{2011ApJ...740..102K}). The formation and evolution of large scale structure has been studied thoroughly and appears to be in remarkable agreement with a diversity of numerical simulations. 
Similar studies of small scale structures encountered discrepancies such as the `missing satellite' problem (\citealt{1993MNRAS.264..201K}; \citealt{1999ApJ...522...82K}; \citealt{1999ApJ...524L..19M}).

The problem refers to the overabundance of predicted Cold Dark Matter sub-halos compared to satellite galaxies observed in the Local Group. The most popular interpretation nowadays to this lack of low mass galaxies is that the least massive sub-halos are likely to host extremely faint galaxies or even to be almost completely dark matter dominated. I.e., the smallest dark matter halos do exist but they are extremely inefficient at forming stars and thus could not create an observable dwarf galaxy.
Physical processes like supernovae feedback, heating from photoionization, the ability for gas to cool and tidal plus ram pressure stripping, can suppress galaxy formation and thus reduce the number of satellites significantly, leaving a population of truly dark sub-halos (\citealt{2008ApJ...679.1260M}; \citealt{2009MNRAS.397L..87L}; \citealt{2010ApJ...710..408B}; \citealt{2009ApJ...692L.109M}; \citealt{2011MNRAS.410.1975W}; \citealt{2011MNRAS.417.1260F}; \citealt{2010AdAst2010E..33R}; \citealt{2014arXiv1412.2748S}). The question from that standpoint is to identify a minimum halo mass that can host a luminous galaxy, where the truncation in the efficiency of galaxy formation occurs.

One of the difficulties in studies of the Local Group is our particular vantage point inside the Milky Way, which means that large completeness corrections need to be applied (\citealt{2008ASSP....5..195K}; \citealt{2008ApJ...688..277T}; \citealt{2009AJ....137..450W}; \citealt{2010ApJ...717.1043B}; \citealt{2014MNRAS.440.3511T}; \citealt{2014ApJ...795L..13H}). In principle, studies of galaxies outside of the Local Group can provide a better characterization of the number of faint galaxies, as well as the galaxy-to-galaxy scatter. Current data on external galaxies do not reach the depths of the Milky Way studies, but with data such as presented here we can make a start in identifying the faintest, most diffuse group members of Milky Way mass galaxies outside of the Local Group.

Figure \ref{fig:clf} points out that the M101 group may suffer from a `missing satellite' problem, similarly to the Local Group. 
Due to the low abundance of satellite galaxies associated with this group, the discrepancy between the observed number of satellites and the predicted number of dark matter sub-halos is similar to that of Milky-Way and M31.
The field of view in this study covers $\sim 82\%$ of the virial volume of the M101 group halo, where a volume incompleteness correction resulted in a $\sim 92\%$ completeness. So even when applying a coverage completeness, there exists a large overabundance of predicted sub-halos compared to satellite galaxies. 
Future wide field photometric surveys are essential for detecting ultra-faint dwarf galaxies and might shed light on the `missing satellite'  problem from both theoretical and observational points of view.

This result might be related to the `missing stellar halo' of M101, presented in \citealt{2014ApJ...782L..24V}. These authors show that the halo mass function of M101, $f_{\text{halo}} = M_{\text{halo}}/M_{\text{tot,*}} = 0.003$, is lower than that of Milky Way ($f_{\text{halo}} \sim 0.02$) and M31($f_{\text{halo}} \sim 0.04$). All three galaxies fall below the $f_{\text{halo}} - M_{\text{tot,*}}$ relation predicted by recent cosmological simulations, with $\text{M101}'\text{s}$ halo mass a factor of ~10 below the median exception. This may fall into line with the even more severe `missing satellite' problem presented in the M101 group, compared to Milky Way and M31.
We note that we cannot rule out the possibility that very large, very low surface brightness galaxies exists in the field.
The dark matter halo properties of the central galaxy may affect the number of satellites associated with different galaxy groups. For example, a lower mass host is more likely to have fewer satellite galaxies (e.g., \citealt{2009ApJ...696.2115I}; \citealt{2012JCAP...12..007P}; \citealt{2015ApJ...810...21M}). Recent works have shown that other properties including mass assembly history, concentration, subhalo population of the host halo, formation redshift of the host halo and baryonic processes, may impact the properties of the satellite galaxy population (e.g., \citealt{2005ApJ...624..505Z}; \citealt{2006ApJ...639L...5Z}; \citealt{2015ApJ...810...21M}; \citealt{2016ApJ...830...59L}; \citealt{2016arXiv161002399J}).

% \citealt{2012JCAP...12..007P};
%Very large galaxies that would be fainter than our surface brightness limit might have total luminosity as galaxies described here but would not have been detected due to their low, out of the limit, surface brightness. 
%Very large galaxies with surface brightness fainter than our survey limit may have a total luminosity as galaxies described here but wouldn't be detected due to their low, out of the limit, surface brightness. Deeper photometric data, in lower surface brightnesses, would provide a better certainty. 

\subsection{A `Too Big to Fail' Problem for M101}\label{tbtf}

\begin{figure}[t]
  \includegraphics[width=\linewidth]{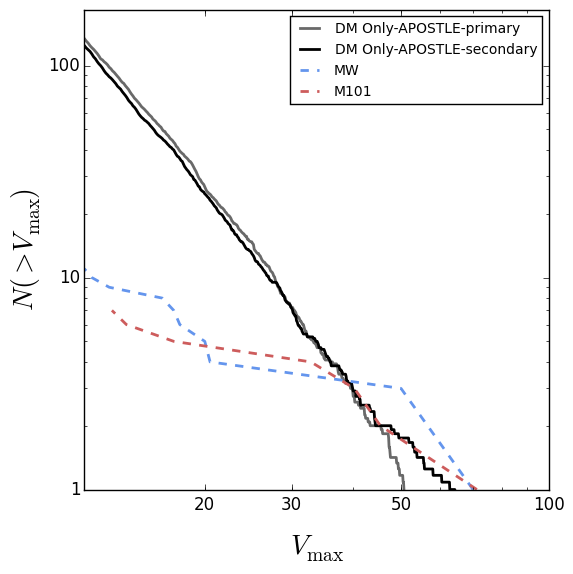}
  \caption{Cumulative number of satellite galaxies above a maximum circular velocity, $V_{\text{max}}$, observed around the Milky Way (blue) and M101 (red). Grey and black curves are from the dark matter only simulation APOSTLE (\citealt{2016MNRAS.457.1931S}).}
  \label{fig:TBTF}
\end{figure}

Another disagreement emerged on smaller scales structure is the `too big to fail' problem, first presented by Boylan-Kolchin et al. (\citeyear{2011MNRAS.415L..40B}, \citeyear{2012MNRAS.422.1203B}).
They have used 6 hydrodynamical simulations, the Aquarius Suite (\citealt{2008MNRAS.391.1685S}), designed to simulate Milky Way dark-matter halos with a variety of mass and force resolution. By comparing the central masses of Milky Way dSph satellites deduced from their kinematics with those of dark matter sub-halos in simulations, they have showed that simulations of Milky Way equivalents contain an order of magnitude denser sub-halos than those compatible with the most luminous observed dSphs. Furthermore, their likelihood analysis of the Aquarius data has predicted that all of the Milky Way dSphs reside in halos with $V_{\text{max}} \sim 25 \text{ km}\text{ s}^{-1}$, whereas more than 10 sub-halos per host halo are expected to have $V_{\text{max}} > 25\text{ km}\text{ s}^{-1}$.

Until recently, `too big to fail' studies have largely dealt with the dark-matter only (DMO) sub-halos and observed satellites in Milky Way. \citealt{2014MNRAS.440.3511T} have shown that the `too big to fail' problem in the Milky Way was not a statistical fluke by showing that M31 exhibits the same problem. \citealt{2014MNRAS.444..222G} showed the `too big to fail' problem exists for isolated dwarf galaxies in the Local Field, beyond the virial radii of the Milky Way and M31, and eliminated the uncertain effects introduced by environment. Recently, \citealt{2015MNRAS.453.3575J} have used semi-analytical models to construct thousands of realizations of Milky Way size host halos and showed `too big to fail' problem with unprecedented statistical power.

 A simple characterization of the `too big to fail' problem is given by the number of satellite halos with maximum circular velocity, $V_{\text{max}}$, above $\approx 30 \text{ km}\text{ s}^{-1}$, where all satellite halos are expected to be luminous (\citealt{2014arXiv1412.2748S}; \citealt{2016MNRAS.457.1931S}). 
Figure \ref{fig:TBTF} shows the cumulative number of satellites above a given $V_{\text{max}}$. Dashed blue and red curves show satellites around the Milky Way (blue) and M101 (red), while the black curve corresponds to the dark matter only simulation, APOSTLE (\citealt{2016MNRAS.457.1931S}). Only three Milky Way satellites are consistent with halos more massive than this limit (the two Magellanic Clouds and the Sagittarius dwarf), whereas DMO simulations of Milky Way sized halos produce two to three times this number. Indeed, as shown in Figure \ref{fig:TBTF}, the DMO halos contain an average of 7-8 satellites with $V_{\text{max}} > 30 \text{ km}\text{ s}^{-1}$, twice than the observed number of luminous M101 satellites and more than twice than the observed number of luminous Milky Way satellites. We therefore cautiously conclude that the M101 satellite's dynamics are consistent with the satellites of the Milky Way. If dark matter mass estimated are obtained from their rotation curves, we predict that the M101 satellites will exhibit the `too big to fail' problem.

\section{Conclusions}
We measured the distances, based on two-color imaging with HST ACS and using the TRGB method, to galaxies in the field of nearby spiral galaxy M101. We have confirmed three new members of the M101 group, out of seven original candidates first discovered with the Dragonfly Telephoto Array. This yields the following main results:

\begin{enumerate}[label=(\roman*)]
\item The M101 group consists of seven satellite galaxies: NGC 5474, NGC 5477, UGC 8837, UGC 9405, DF1, DF2 and DF3. Compared to the Milky Way and M31 groups, these galaxies span a wider range of luminosities, $6 \lesssim \log(L/L_{\odot}) \lesssim 9.5$.

\item For $M_V \lesssim -15$ the observed cumulative luminosity functions of satellites around the Milky-Way, M31 and M101 are remarkably similar. For fainter magnitudes, down to $M_V = -9.2 \text{ mag}$, we find only three satellites of M101. 
%Their low abundance is an evidence a `missing satellite' problem in the M101 group, outside the Local Group. This result may be related to the `missing stellar halo' of M101, presented in \citealt{2014ApJ...782L..24V}.
The low number of satellites may be evidence for a `missing satellite' problem in the M101 group, outside the Local Group. This result may be related to the `missing stellar halo' of M101, presented in \citealt{2014ApJ...782L..24V}.
The main observational caveat is the completeness of the sample in terms of the areal coverage of the field around M101 and the limiting absolute magnitude. The present study can be expanded to a wider field around M101 and longer integration times would result in lower surface brightness limits. 

\item Assuming the $L/L_{\odot}-V_{\text{max}}$ relation for the M101 satellites is similar to that in the Local Group, the lack of intermediate mass galaxies in the M101 group and the previously measured values for three out of seven group members suggests that M101 exhibits a similar `too big to fail' problem as the Milky Way and M31.

%\item The M101 satellites' dynamics are predicted to be consistent with those of the satellites of the Milky Way and M31, and exhibit a `too big to fail' problem. 
%A future study that will compare between the kinematics of the M101 satellite galaxies and a sample of subhalos from a plausible choice of an M101-like halo is necessary in order to confirm our cautiously stated inference.

\end{enumerate}

%This M101 group study was the first Dragonfly discovered-HST followed type study. 
More low surface brightness objects have been discovered in four Dragonfly fields, centered on NGC 1052, NGC 1084, NGC 3384 and NGC 4258 and these have been scheduled for HST ACS observations. Constructing the satellite luminosity function in these four fields will provide a better understanding of whether other small-scales faults presented by $\Lambda\textrm{CDM}$ hold for a larger, statistical sample of galaxies beyond the Local Group.

\acknowledgments
\section*{Acknowledgement}
We thank the anonymous referee for a helpful and constructive report. Support from STScI grant HST-GO-13682 is gratefully acknowledged.
We thank Marla Geha (PI of the SAGA Survey) and Erik Tollerud for valuable discussions. We also wish to thank Andrew Dolphin for his help with DOLPHOT, Till Sawala and the APOSTLE team for providing access to their simulation data and Dmitry Makarov for providing access to the TRGBTOOL used in this work. Lastly, we thank Chen Zirinski for the graphics assistance. 
IDK and LNM acknowledge the support of the Russian Science Foundation grant 14-12-00965.
\bibliography{paper1_ref}

\end{document}